\documentclass[fleqn,usenatbib]{mnras}

\usepackage{newtxtext,newtxmath}

\usepackage[T1]{fontenc}
\usepackage{ae,aecompl}

\usepackage{graphicx}	
\usepackage{amsmath}	

\title[AGN variability model]{Exploring black-hole scaling relations via the ensemble variability of Active Galactic Nuclei}

\author[Georgakakis et al.]{
A. Georgakakis$^{1}$\thanks{E-mail: age@noa.gr}, 
I. Papadakis$^{2,3}$,
M. Paolillo$^{4,5,6}$
\\
$^{1}$Royal Astronomical Society, Burlington House, Piccadilly, London W1J 0BQ, UK\\
$^{2}$Department of Physics and Institute of Theoretical and Computational Physics, University of Crete, 71003 Heraklion, Greece\\
$^{3}$Institute of Astrophysics, Foundation for Research and Technology, 71110 Heraklion, Greece\\
$^4$Dipartimento di Fisica "Ettore Pancini", Universit\`a Federico II, via Cinthia, 80126 ,Napoli, Italy\\
$^5$INAF - Osservatorio Astronomico di Capodimonte, via Moiariello 16, 80131, Napoli\\
$^6$Istituto Nazionale di Fisica Nucleare, Sezione di Napoli, I-80126 Napoli, Italy
}

\date{Accepted XXX. Received YYY; in original form ZZZ}

\pubyear{2015}

\begin{document}
\label{firstpage}
\pagerange{\pageref{firstpage}--\pageref{lastpage}}
\maketitle

\begin{abstract}
An empirical model is presented that links, for the first time, the demographics of AGN to their ensemble X-ray variability properties. Observations on the incidence of AGN in galaxies are combined with (i) models of the Power Spectrum Density (PSD) of the flux variations of AGN and (ii) parameterisations of the black-hole mass vs stellar-mass scaling relation, to predict the mean excess variance of active black-hole populations in cosmological volumes. We show that the comparison of the model with observational measurements of the ensemble excess variance as a function of X-ray luminosity provides a handle on both the PSD models and the black-hole mass vs stellar mass relation. We find strong evidence against a PSD model that is described by a broken power-law and a constant overall normalisation. Instead our analysis indicates that the amplitude of the PSD depends on the physical properties of the accretion events, such as the Eddington ratio and/or the black hole mass. We also find that current observational measurements of the ensemble excess variance are consistent with the black-hole mass vs stellar mass relation of local spheroids based on dynamically determined black-hole masses. We also discuss future prospects of the proposed approach to jointly constrain the PSD of AGN and the black-hole mass vs stellar mass relation as a function of redshift. 
\end{abstract}




\section{Introduction}

One of the fundamental properties of the accretion flows onto supermassive black holes (SMBHs) is the variability of the radiated flux. Such stochastic variations occur on a wide range of timescales and provide information on the size of the central source \citep[e.g.][]{Lynden-Bell1969} and the physics of the accretion process \citep[e.g.][]{Rees1984}. The origin of these variations is still under discussion and could be related to instabilities of the accretion flow, a flaring corona or hotspots orbiting the central compact object \citep[e.g.][]{GravityCollaboration2018, GravityCollaboration2020}. Whatever the nature of the underlying physical mechanism, observations, particularly at X-rays, point to a common process for the flux variability of active black holes over a broad range of masses and accretion rates \citep[e.g.][]{McHardy2006, Koerding2007}. This is manifested by remarkable similarities in the statistical measures of the observed flux variations (e.g. Power Spectral Density) of different objects, once the key physical parameters of individual systems, such as the mass of the compact object and/or the Eddington ratio of the accretion flow, are factored out \cite[e.g.][]{Gonzalez-Martin_Vaughan2012, Ponti2012}. These similarities extend from supermassive black-holes in AGN to stellar-size black holes in binary systems, thereby indicating common variability mechanisms over many orders of magnitude in mass. The implication of this observational fact is that the amplitude of the variability on different timescales provides a handle on the  physical properties of the accreting system. For example, flux variations, particularly at X-rays, have being proposed as a means of measuring the black-hole masses of AGN \citep[e.g.][]{Czerny2001, Nikolajuk2004, Ponti2012} in a way that is complementary to dynamical estimates. 

In addition to studies of the light curves of individual objects, it has also been shown that there is value in measuring the mean variability properties of AGN populations detected in extragalactic X-ray survey fields \citep{Paolillo2004, Papadakis2008, Paolillo2017}. These measurements are taking advantage of the fact that in many popular survey fields the total integration time has been gradually built up by numerous repeat observations carried out over the course of many years \citep[e.g. 7\,Msc Chandra Deep Field South,][]{Luo2017}. Although the light curves of individual sources in such surveys carry limited information, the ensemble of all AGN provides useful constraints on the integrated variability power of the population \citep{Allevato2013_var}. Such observations have enabled investigations on the redshift evolution of the AGN variability properties \citep[e.g.][]{Papadakis2008} and the dependence of the flux variability amplitude on observables such as the accretion luminosity \citep[e.g.][]{Paolillo2017}. Moreover, because of the dependence of the variability power spectrum on the physical properties of the active black holes, measurements of the ensemble flux variations of AGN contain information on the distribution of Eddington ratios and black-hole masses of the population \citep[e.g.][]{Allevato2010, Paolillo2017}. 

The latter two quantities are also relevant to investigations of the accretion history of the Universe and the co-evolution of AGN and their host galaxies. Observational studies on the incidence of AGN in galaxies for example, associate proxies of the Eddington-ratio distribution to the properties of AGN hosts (e.g. star-formation rate, stellar mass) to explore the physical conditions that promote accretion events onto SMBHs \citep[e.g.][]{Kauffmann_Heckman2009, Azadi2015, Georgakakis2017_plz, Aird2018, Aird2019}. Also, the  continuity equation of the black-hole mass function uses the observed AGN luminosity function as boundary condition to determine the growth history of SMBHs,  infer the Eddington-ratio distribution of AGN and constrain black-hole fuelling models \citep[e.g.][]{Merloni_Heinz2008, Shankar2013, Aversa2015}. Observational measurements of the ensemble variability of AGN could feedback to the studies above by providing an independent observational handle on the Eddington ratio and black-hole mass distributions of the population. 

In this paper we build upon this potential to link the mean variability of AGN populations to black-hole demographics. A new empirical model is developed that combines observational results on the occupation of galaxies by AGN with models of their variability amplitude on different timescales. This is used to make predictions on the mean variability of AGN populations as a function of observables, such as accretion luminosity and redshift. A forward modeling approach is then used to compare the predictions with observations.  We demonstrate the predictive power of the model and show how it can jointly constrain models of the AGN variability amplitude and the black-hole mass vs stellar mass relation of the population.

\section{Model Construction}\label{sec:model}

\begin{figure*}
\begin{center}
\includegraphics[width=2.0\columnwidth]{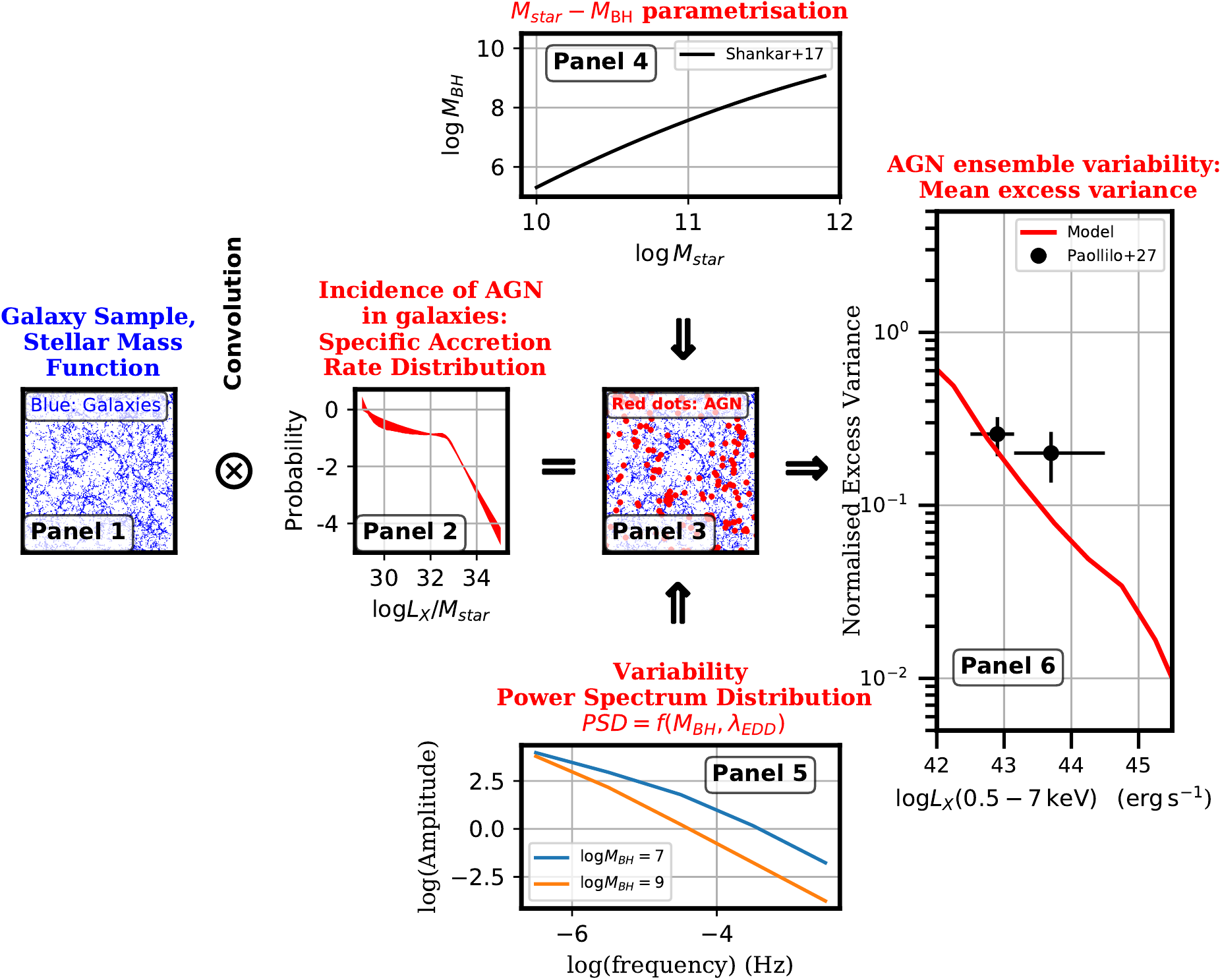}
\end{center}
\caption{Flowchart of the empirical AGN variability model. The blue dots in Panel 1 represent galaxies distributed in a cosmological volume that are drawn from the stellar mass function \protect\citep{Ilbert2013}. They are seeded with AGN specific accretion-rates ($\lambda \propto L_X / M_\star$) using the corresponding observationally-derived probability distribution functions, e.g.  those shown in Panel 2. This produces a sample of mock AGN (red dots of Panel 3), each of which has been assigned an X-ray luminosity, $L_X$, a host-galaxy stellar mass, $M_\star$, and a redshift, $z$. A parametrisation of the $M_\star-M_{BH}$ correlation (Panel 4) is used to assign black holes to mock AGN and hence, Eddington ratios $\lambda_{EDD} \propto L_{X} / M_{BH}$. The dependence of the AGN variability Power Spectrum Density (PSD; Panel 5) on Black Hole Mass and Eddington ratio is then used to assign variability to each mock AGN. It is then straightforward to determine for each AGN an excess variance  ($\sigma_{NXV}^2$) by integrating the corresponding PSD. The average excess variance of the population binned in luminosi ty and redshift intervals (solid line in Panel 6)  can then be directly compared with observational results \protect\citep[black points in Panel 6;][]{Paolillo2017}.
}\label{fig:workflow}
\end{figure*}

  
The construction of the AGN ensemble-variability model is based solely on empirical (i.e. observationally-derived) relations.  The workflow is graphically demonstrated in Figure \ref{fig:workflow}. The starting point are recent observationally-determined AGN specific accretion-rate distributions, $P(\lambda)$  \citep[Panel-2 of Fig.-\ref{fig:workflow}; e.g.][]{Georgakakis2017_plz, Aird2018}, which measure the probability of a galaxy hosting an accretion event with specific accretion-rate $\lambda\propto L_{X}/M_{\star}$. In this definition $L_X$ is the X-ray luminosity of the event (AGN) in a given spectral band and $M_\star$ is the stellar mass of the host galaxy. The specific accretion-rate is a purely observationally-derived parameter and measures how much X-rays an AGN emits relative to the stellar mass of its host galaxy. The feature of the specific accretion-rate distribution, $P(\lambda)$, is that it is a measure of the incidence of AGN among galaxies. Therefore, it can be applied in a probabilistic way to a galaxy sample (see Panel-1/Fig.-\ref{fig:workflow}) and seed them with AGN (see Panel-3/Fig-\ref{fig:workflow}). Mathematically, this seeding process is described by the convolution of the stellar mass function of galaxies \citep[a well-constrained observable, e.g.][]{Weigel2016} with the specific accretion-rate distribution. The resulting mock-AGN sample is consistent, by construction, with the evolving X-ray luminosity function of AGN. This modeling approach also successfully reproduces the observed stellar-mass function of AGN hosts \citep{Georgakakis2017_plz}, the distribution of AGN on the cosmic web \citep{Leauthaud2015, Georgakakis2019, Aird_Coil2020} and their multi-wavelength properties \citep{Georgakakis2020}. The end-products of the process described above is a mock sample of AGN, each of which is assigned an X-ray luminosity, host galaxy stellar mass and redshift ($L_X$, $M_{\star}$, $z$). This parameter space is expanded to include the black-hole mass of individual systems. First a parametrization of the black-hole mass vs stellar mass relation is introduced (panel 4 of Fig. \ref{fig:workflow}). Each mock AGN with stellar mass, $M_\star$, is then assigned a black-hole mass, $M_{\rm BH}$ and given the X-ray luminosity, $L_X$, an Eddington ratio $\lambda_{EDD} \propto L_X/L_{EDD}$, where $L_{EDD}$ is the Eddington luminosity. Next we describe how this extended parameter space ($L_X$, $M_{\star}$, $z$,  $M_{\rm BH}$) is used to model the variability amplitude of the individual AGN and the ensemble.

The focus of this work is the modeling of the stochastic variations of the luminosity of AGN that occur on different timescales. They are quantitatively described by the Power Spectral Density (PSD, Panel 5 of Fig. \ref{fig:workflow}), which describes the distribution of the light-curve variance in Fourier frequencies. Observations show that the PSD of nearby Seyferts can be approximated with a broken power-law functional form with parameters (i.e slopes, break frequency, normalisation) that depend on the physical properties of the system, such as the mass of the black hole and the accretion rate onto it \citep[e.g.][]{McHardy2006, Kording2007, Ponti2012}. These observations point to a common physical mechanism for the aperiodic flux variations of  AGN and indicate that the observed variability is coupled to the physical parameters of the active black hole. Observationally, the measurement of the PSD of AGN requires long and interrupted monitoring campaigns  that are currently available for only a few dozen systems. An alternative approach for studying the variability of large numbers of active black holes is the  normalised excess variance \citep[$\sigma^2_{\rm NXV}$,][]{Nandra1997}. This is less demanding on resources and provides an estimate of the integral of the PSD over the timescales of the observations. The  normalised excess variance is the quantity we choose to use to compare the model predictions against observations. Following  \cite{Paolillo2017}  we adopt analytical empirical relations that link the PSD parameters to the black-hole mass and the Eddington ratio of the accretion event. The $M_{\rm  BH}$ and $\lambda_{EDD}$ of the mock AGN are then plugged into these relations to  compute their $\sigma^2_{\rm NXV}$.

We are interested in the ensemble (mean) excess variance of the AGN population rather than the variability properties of individual systems. The Panel-6 of Figure \ref{fig:workflow} shows the parameter space that will be used in later sections to compare the model predictions against the observations. It plots the ensemble variance of AGN as a function of X-ray luminosity. The data points on this plot are measurements of the mean $\sigma^2_{NXV}$ of AGN in the Chandra Deep Field South field \citep{Paolillo2017}. On the model side, given a population of mock AGN each of which is assigned a PSD, it is possible to estimate the excess variance of individual sources. The sample can then be binned by X-ray luminosity to yield the mean (ensemble) excess variance as a function of $L_X$ and compare directly with the observations as shown in Panel-6 of Fig. \ref{fig:workflow}. In the next sections we describe in detail each of the components of the ensemble variability model that is  graphically demonstrated in the Fig. \ref{fig:workflow}.

\subsection{The Stellar Mass Function }\label{sec:SMF}

For the stellar mass function of galaxies we adopt the double Schechter-function parametrisation presented by \cite{Ilbert2013} based on observations in the COSMOS survey field \citep{Scoville2007}. They provide analytic fits to the galaxy mass function in discrete redshift intervals between $z=0.2$ and $=4$. These are interpolated/extrapolated to yield a continuous sampling of the mass function in the redshift range $z=0-4$. Below the redshift limit of $z=0.2$ the mass function is fixed to the parametrisation of the lower redshift bin of \cite{Ilbert2013}. The resampled mass functions define a 2-dimensional surface in the stellar mass vs redshift space. This is used to randomly draw pairs of $M_\star$ and $z$ that are distributed in the above 2-dimensional space according to the observations. This sequence of pairs represents the mock galaxy sample. 

\subsection{Specific Accretion Rate Distribution}\label{sec:SARD}

Mock galaxies are seeded with specific accretion rates $\lambda$, using the probability density distributions, $P(\lambda)$, presented by \cite{Georgakakis2020}. The latter are approximated by a broken power-law with parameters determined by requiring that the convolution of the $P(\lambda)$ with the galaxy stellar mass function of \cite{Ilbert2013} yields the total X-ray luminosity function measured by \cite{Aird2015}. Each mock  galaxy with stellar mass $M_{\star}$ is assigned a specific accretion rate, $\lambda \propto L_X / M_\star$, which is drawn from the distributions presented by \cite{Georgakakis2020}. The intrinsic (i.e. corrected for obscuration) X-ray luminosity of a given mock AGN is estimated as $L_X = \lambda \times M_\star$, where in this application the $L_X$ corresponds to the 2-10\,keV spectral band. 

In the following sections the X-ray fluxes of mock AGN will  also be required. This is to mimic the observational selection effects of flux-limited AGN samples and provide a meaningful comparison between the model predictions with the observations. The determination of model fluxes requires knowledge of the level obscuration of individual mock AGN that absorbs their intrinsic luminosities. An X-ray spectral model is also needed to convert luminosities to fluxes. 

Obscured AGN are accounted for in the estimation of the X-ray luminosity function of \cite{Aird2015} and are therefore included in our modeling. The AGN obscuration is parameterized by the atomic-hydrogen column density, $N_H$. The distribution of AGN in $N_H$ is a function of both accretion luminosity and redshift following the model presented by \cite{Aird2015}. Compton thick AGN with $N_H>10^{24}\rm \,cm^{-2}$ are also included in this model. Their space density is assumed to be 34\% of moderately obscured active black holes, i.e. those with $N_H=10^{22}-10^{24}\rm \,cm^{-2}$. The \cite{Aird2015} model distribution is sampled in a probabilistic way using a Monte Carlo approach to assign mock AGN  line-of-sight atomic-hydrogen column densities. 

Using the $N_H$, $z$, $L_X(\rm 2-10\,keV)$ assigned to mock AGN it is then also possible to estimate the corresponding flux in any observed energy band. This calculation follows the methodology described in \cite{Georgakakis2020}. The adopted X-ray spectrum consists of an intrinsic power-law that is transmitted through an obscuring medium that absorbs and scatters the X-ray photons. We use the torus model of \cite{Brightman_Nandra2011} to describe  these  processes and  produce  the  resulting  X-ray  spectra. This model assumes a sphere of constant density with two symmetric conical wedges with vertices at the centre of the sphere removed. The opening angle of the cones is fixed to 45 degrees and the viewing angle of the observer is set to 87 degrees, i.e. nearly edge on. The spectral index of the intrinsic power-law is assumed to be $\Gamma=1.9$ \citep[e.g.][]{Nandra1997}. 

\subsection{Relation between stellar and black-hole mass}\label{sec:MSBH}

Two different parameterizations of the black-hole vs stellar mass relation are adopted. The first is based on dynamically measured black-hole masses at the centres of local non-active galaxies, i.e. those with dormant black holes. We use the scaling relation 

\begin{equation}\label{eq:MstarMBH_dyn} 
    \log \frac{M_{\rm BH}}{M_\odot} = 8.35 + 1.31 \, \left( \log \frac{M_{\star}}{M_\odot} - 11 \right),
\end{equation}

\noindent which is derived by \cite{Shankar2020Nat} based on the sample of early and late-type galaxies with dynamical black-hole mass estimates presented by \cite{Savorgnan2016}. The coefficients of the relation above are estimated using all the galaxies in the sample of  \cite{Savorgnan2016}. The intrinsic rms scatter in the $\log M_{\rm BH}$-direction is 0.5\,dex.

Evidence has been emerging recently suggesting that the scaling relations based on dynamical black-hole mass estimates, like the one in Equation \ref{eq:MstarMBH_dyn}, may be biased \citep{Bernardi2007, Shankar2016}. This is because the gravitational sphere of influence of supermassive black holes has to be resolved to estimate their masses via dynamical arguments. At the spatial resolution limit of current instrumentation this is feasible only for the subset of local galaxies that host the most massive black holes at fixed stellar mass \citep{Shankar2016}. It is argued that this selection effect distorts the inferred normalization and/or shape of the local scaling relations between stellar mass proxies and black-hole masses \citep{Shankar2017}. We explore the impact of this potential source of bias on the modeling of the AGN variability by also considering the intrinsic (unbiased) scaling relation proposed by \cite{Shankar2016} 

\begin{equation}\label{eq:MstarMBH_int}
\begin{split}
    \log \frac{M_{\rm BH}}{M_\odot} = & 7.574 + 1.946\,\left( \log \frac{M_{\star}}{M_\odot}  -11 \right)\\
    & -0.306\,\left( \log \frac{M_{\star}}{M_\odot} -11 \right)^2\\
    &-0.011\,\left( \log \frac{M_{\star}}{M_\odot} -11 \right)^3.
\end{split}    
\end{equation}

\noindent In the relation above the black-hole mass logarithmic scatter is assumed to depend on $M_{\star}$ as in \cite{Shankar2016} 

\begin{equation}\label{eq:MstarMBH_sc}
    \sigma_{\rm BH} = 0.32 -0.1 \, \left( \log \frac{M_{\star}}{M_\odot}  -12 \right).
\end{equation}

\noindent It is cautioned that the stellar mass in the relations above should represent that of the bulge, $M_{\rm bulge}$, not the total of the galaxy. Equations \ref{eq:MstarMBH_dyn}, \ref{eq:MstarMBH_int} therefore assume that $M_{\star} \approx M_{\rm bulge}$. This approximation breaks down in the case of late-type and/or bulgeless galaxies. As an example, spiral galaxies of the Sb/Sbc type have typical bulge to total stellar mass ratios $M_{{\rm bulge}}/M_{\star} \approx 0.5$ \citep[e.g][]{Fukugita1998, Oohama2009}. The impact of this effect on the results will be investigated in later sections. 

Equations \ref{eq:MstarMBH_dyn}, \ref{eq:MstarMBH_int} will be used independently to seed galaxies with black holes and produce distinct predictions on the variability amplitude of the resulting samples. For a given black hole mass the corresponding Eddington ratio is $\lambda_{Edd} = L_{bol}/L_{Edd}$, where $L_{Edd}$ is the Eddington luminosity. The bolometric luminosity, $L_{bol}$, is estimated from the 2-10\,keV X-ray luminosity using the bolometric correction of \cite{Duras2020} for their combined Type-1 and Type-2 AGN sample.

\subsection{Variability model parameterisation}\label{sec:psd}

The focus of this work are the aperiodic time variations of the AGN flux. These can be characterised by the PSD that describes how the variability amplitude is distributed in Fourier frequencies. X-ray monitoring campaigns of a few dozen luminous AGN show that their PSDs can be approximated with a double power-law functional form with a slope of about --2 at high frequencies that flattens to  --1 at the low frequency end \citep[e.g.][]{Papadakis2002, Uttley2002, Markowitz2003, McHardy2007}. We parametrise the PSD with a bending power-law of the form

\begin{equation}\label{eq:psd}
PSD(\nu) = A\,\nu^{-1} \, \left(1+    \frac{\nu}{\nu_b}\right) ^{-1},
\end{equation}

\noindent similar to that proposed by \citet[][see also \citealt{Gonzalez-Martin_Vaughan2012}]{McHardy2004} based on local AGN observations. In the equation above, $A$ is the normalization factor and $\nu_b$ is the bending frequency, where the power-law slope changes from $-1$ at the limit $\nu << \nu_b$ to $-2$ for $\nu >>\nu_b$. The equation above can be integrated to yield the flux variance 

\begin{equation}\label{eq:s2mod}
\begin{split}
\sigma^2_{mod} = & \int_{\nu_{min}}^{\nu_{max}} PSD(\nu)\,d\nu \\ 
  = & A\,\left(\ln\frac{\nu_{max}}{\nu_{min}} - \ln\frac{\nu_b+\nu_{max}}{\nu_b +\nu_{min}}\right),
\end{split}
\end{equation}

\noindent where the integration limits $\nu_{\rm min}$, $\nu_{\rm max}$ are the lowest and highest rest-frame frequencies sampled by the observed light-curve.  These are estimated from the relations

\begin{equation}\label{eq:numin}
    \nu_{min} = \frac{1+z}{\Delta t^{obs}_{max}},
\end{equation}

\begin{equation}\label{eq:numax}
    \nu_{max} = \frac{1+z}{\Delta t^{obs}_{min}},
\end{equation}

\noindent where $z$ is the redshift and $\Delta t^{obs}_{min}$, $\Delta t^{obs}_{max}$ are respectively, the minimum sampled timescale and the total duration of the light curve at the observer's frame.

The parameters of the PSD function of Equation \ref{eq:psd} (i.e $A$, $\nu_b$) are linked to the physical properties of AGN, such as the black-hole mass and the Eddington ratio, using the four observationally motivated models proposed by \cite{Paolillo2017}. 

In the first model (Model 1) the PSD amplitude is constant for all AGN 

\begin{equation}
A = 2\cdot\nu_b \cdot PSD(\nu_b) = 0.02.
\end{equation}

\noindent and the  break frequency scales with the mass of the black hole as

\begin{equation}\label{eq:model1-nub}
\nu_b = \frac{580}{M_{\rm BH}/M_\odot} \; (s^{-1}).
\end{equation}

\noindent These assumptions are based on the observational results of  \cite{Papadakis2004} and  \cite{Gonzalez-Martin_Vaughan2012}.

The second model (Model 2) also assumes a constant PSD amplitude as in Model 1, but the break frequency depends on both the black-hole mass and the accretion rate as proposed by \cite{McHardy2006}. This dependence is expressed in terms of the AGN bolometric luminosity

\begin{equation}\label{eq:model2-nub}
\nu_b=\frac{200}{86400} \cdot \frac{L_{bol}}{10^{44}}\cdot \left(\frac{M_{\rm BH}}{10^6\,M_\odot}\right)^{-2}  \; (s^{-1}),
\end{equation}

\noindent where $L_{bol}$ is the bolometric luminosity in units of $\rm erg\,s^{-1}$ and the black hole mass is measured in solar units. 

The third Model (Model 3) is a variation of Model 1 in that the assumption of a constant PSD amplitude is relaxed. Following the observational results of \cite{Ponti2012} the amplitude is assumed to scale with the Eddington ratio, $\lambda_{Edd}$, of the accretion flow as

\begin{equation}\label{eq:model3-amp}
A = 2\cdot\nu_b \cdot PSD(\nu_b) = 3\times10^{-2}\cdot\lambda_{Edd}^{-0.8}.
\end{equation}

\noindent The break frequency of Model 3 depends on black hole mass as in Equation \ref{eq:model1-nub}.

Finally the fourth model (Model 4) is a mix of Models 2 and 3. The break frequency scales with black-hole mass as in Equation \ref{eq:model2-nub} and the PSD normalization depends on Eddington ratio via Equation \ref{eq:model3-amp}. 

\cite{Allevato2013_var} showed that the normalised excess variance measured from AGN light curves with uneven and/or sparse sampling is not a direct measure of $\sigma^2_{mod}$ as defined in Equation \ref{eq:s2mod} in the case of  PSDs given by Equation \ref{eq:psd}. Instead the normalised excess variance is an estimator of the quantity $\sigma^2_{obs}$  defined as 

\begin{equation}\label{eq:sampling}
\sigma^2_{obs} = \frac{ \sigma^2_{mod} } { C \cdot 0.48^{\beta - 1}}.
\end{equation}

\noindent The parameter $\beta$ depends on the PSD slope below $\nu_{min}$ and $C$ is a correction factor that depends on the sampling pattern. Equation \ref{eq:s2mod} is used to determine the $\sigma^2_{mod}$ of mock AGN. This is then plugged into Equation \ref{eq:sampling} to estimate $\sigma^2_{obs}$, which is used to compare against the observational results. 

\begin{figure}
\begin{center}
\includegraphics[width=0.9\columnwidth]{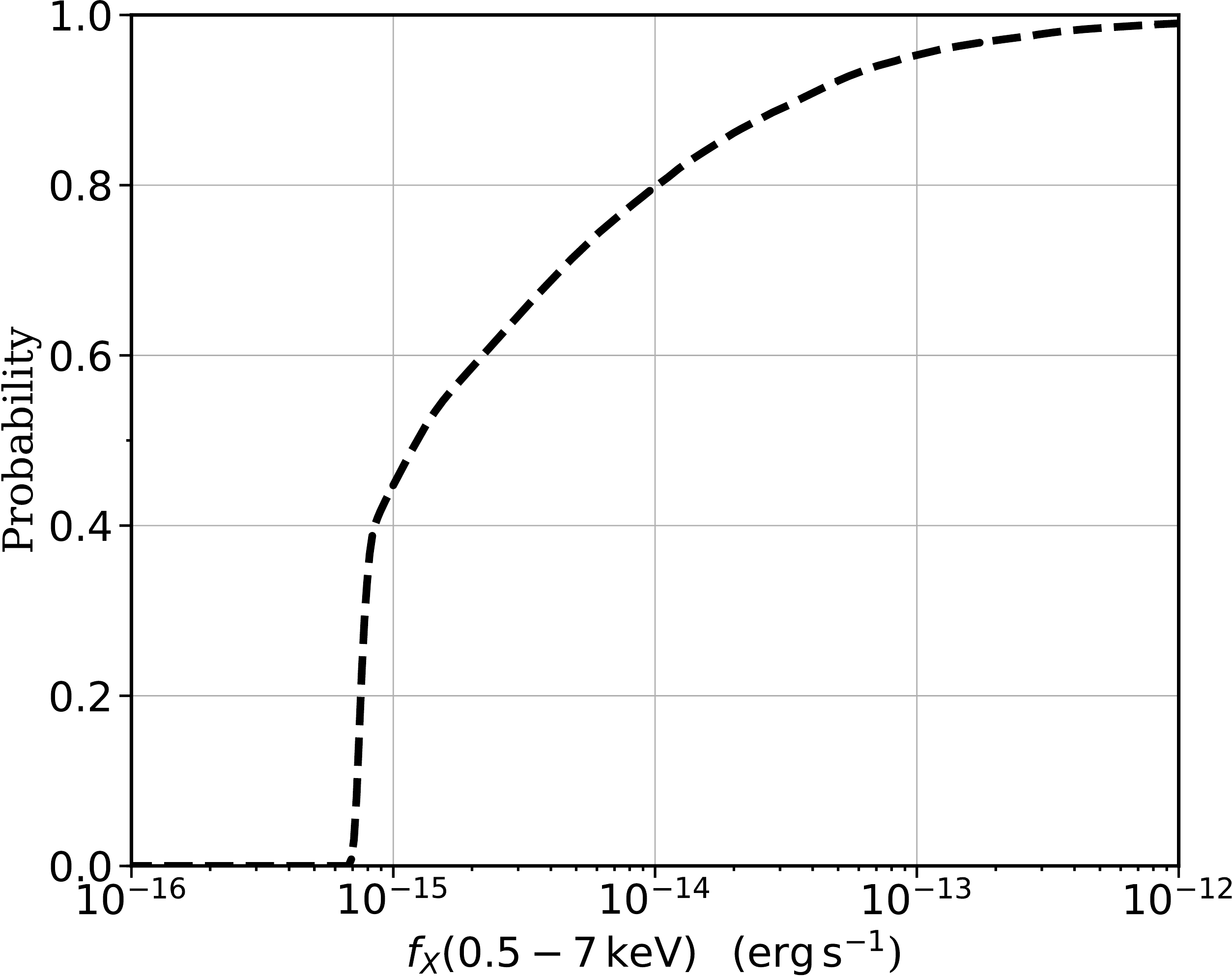}
\end{center}
\caption{The selection function applied to the mock AGN sample.  The vertical axis is the probability of an X-ray source with a given 0.5-7\,keV flux to have more than 350 net counts within the CDFS field of view. This probability is plotted as a function of the 0.5-7\,keV flux on the horizontal axis. The curve is designed to mimic the selection of the \protect\cite{Paolillo2017} CDFS variability sample.}\label{fig:selfun}
\end{figure}

\begin{figure}
\begin{center}
\includegraphics[width=0.9\columnwidth]{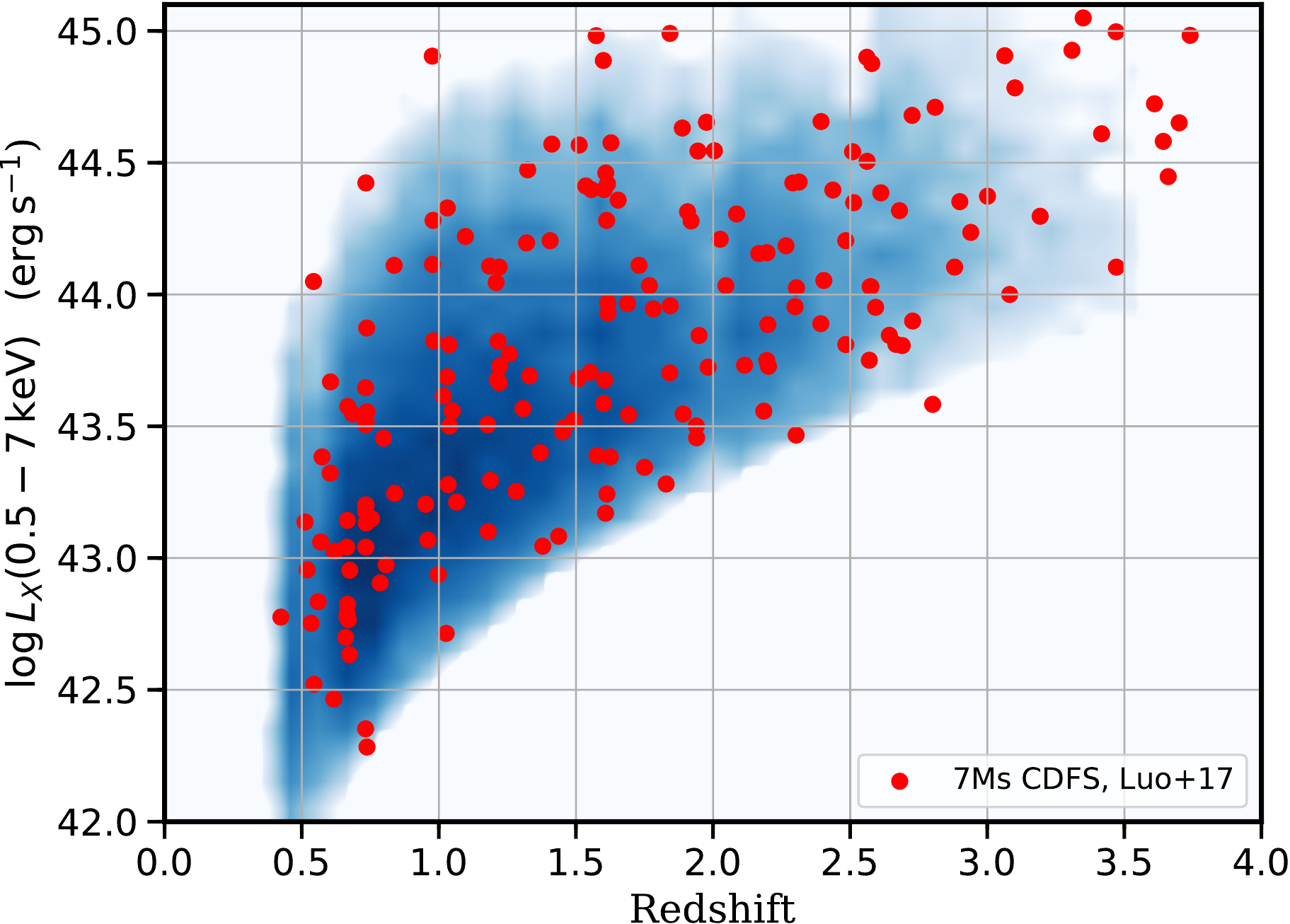}
\end{center}
\caption{X-ray luminosity vs redshift parameter space. The red data points are the X-ray sources in the 7\,Ms CDFS catalogue of \protect\cite{Luo2017} in the redshift interval $z=0.4-4$ and with full-band net counts $>350$. These criteria mimic the selection function of the \protect\cite{Paolillo2017} variability sample. The blue shaded region corresponds to mock AGN in the redshift interval above and filtered through the selection function curve plotted in Fig. \protect\ref{fig:selfun}. Darker shades of blue correspond to a higher density of mock AGN.}\label{fig:lgxz2D}
\end{figure}

\begin{figure}
\begin{center}
\includegraphics[width=0.9\columnwidth]{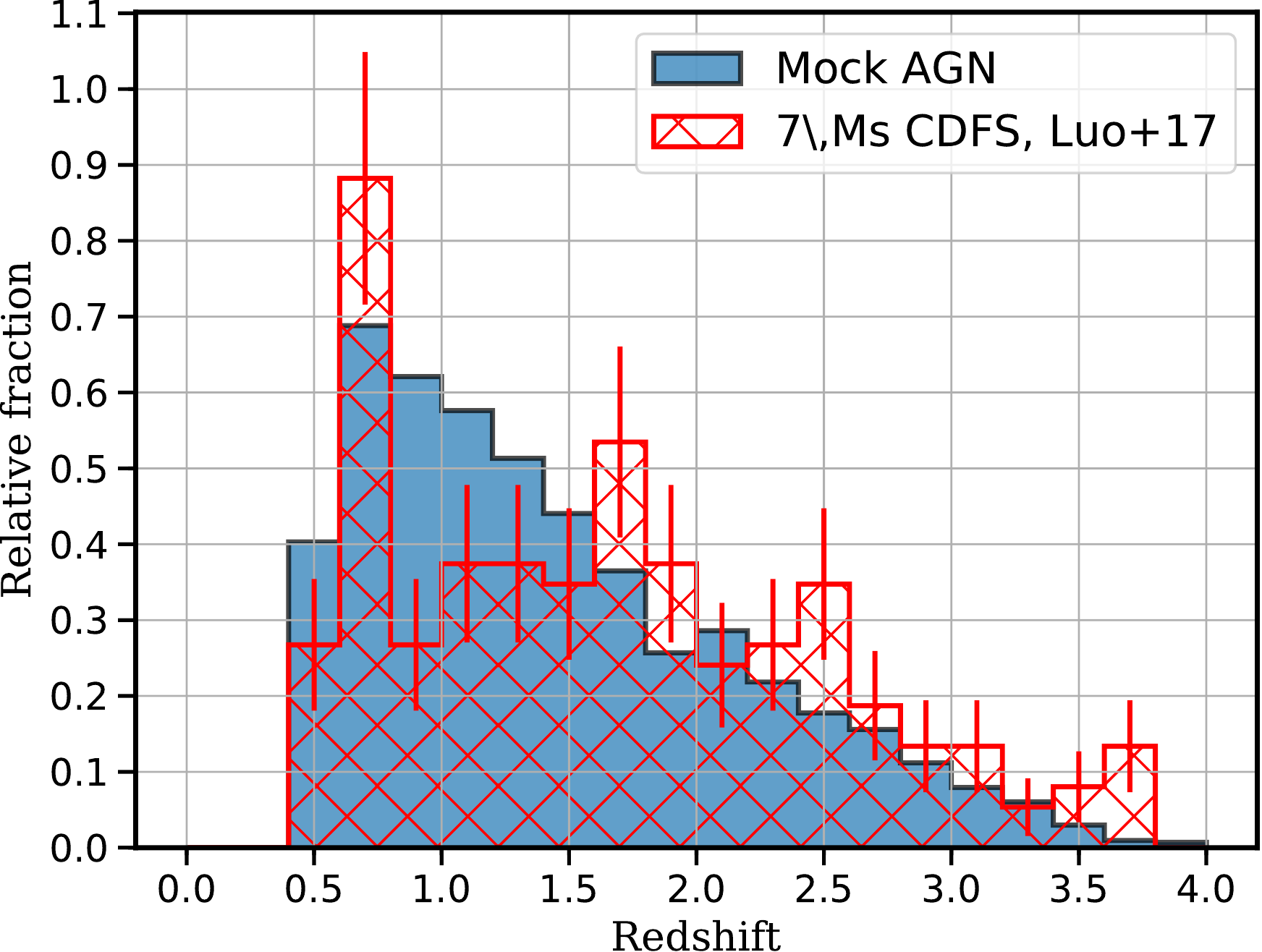}
\end{center}
\caption{The redshift distribution of the 7\,Ms CDFS sources (red hatched histogram) shown in Fig. \ref{fig:lgxz2D}. The errorbars of individual bins correspond to the Poisson uncertainty. The blue histogram is the projection of the 2-dimensional mock AGN distribution of  Fig. \ref{fig:lgxz2D} onto the redshift  axis. Both the blue and red-hatched histograms are normalised to unity.}\label{fig:histz}
\end{figure}

\begin{figure}
\begin{center}
\includegraphics[width=0.9\columnwidth]{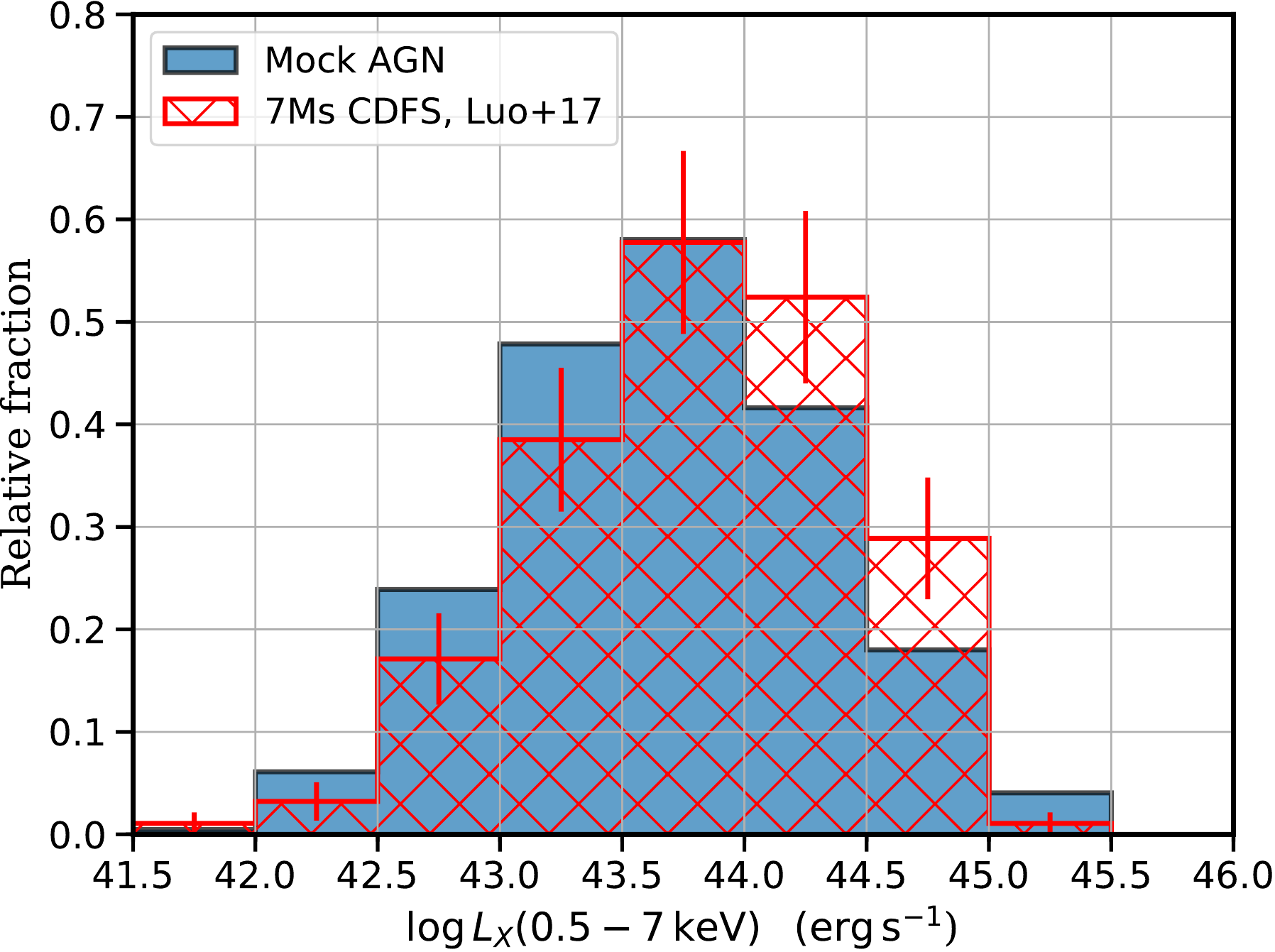}
\end{center}
\caption{The X-ray luminosity distribution of the 7\,Ms CDFS sources (red hatched histogram) shown in Fig. \ref{fig:lgxz2D}. The errorbars of individual bins correspond to the Poisson uncertainty. The blue histogram is the projection of the 2-dimensional mock AGN distribution of  Fig. \ref{fig:lgxz2D} onto the X-ray luminosity axis. Both the blue and red-hatched histograms are normalised to unity. }\label{fig:histlglx}
\end{figure}

\section{Modeling the Observations}\label{sec:observations}

\subsection{The Chandra Deep Field South dataset}

In this paper we use the observational measurements of the AGN ensemble excess variance in the 7\,Ms Chandra Deep Field South \citep[CDFS,][]{Luo2017} presented by \cite{Paolillo2017}. This dataset consists of 102 individual Chandra pointings split into multiple epochs over a period of 17\,years. It currently represents the state-of-the-art in temporal studies of AGN populations because of the large number of repeat observations and their long time-span. 

We compare the model predictions with the excess variance measurements of the full sample of \cite{Paolillo2017} grouped into a single broad redshift bin, $z=0.4-4.0$. They also presented variability measurements in narrower redshift intervals, $z=0.4-1.03$, $1.03-1.8$, $1.8-2.75$ and $2.75-4$. We choose not to use these subsamples because of the larger uncertainties of individual data-points and the narrower luminosity baseline. Additionally, the adopted modeling methodology allows to properly account for the sources' redshift distribution without the need to introduce binning. The excess variance of individual sources is measured from the light-curves that span a timescale of 6205\,days and include all the epochs of the 7\,Ms CDFS observations.

The CDFS variability measurements of \cite{Paolillo2017} are limited to CDFS sources with signal-to-noise ratio $>0.8$. This is nearly equivalent to selecting sources with $>350$ net counts in the $0.5-7$\,keV spectral band of the coadded CDFS observations. For fainter sources the Poisson noise dominates over the intrinsic variability. The thresholds above refer to the photon counts extracted within an aperture of variable size across the CDFS field of view that roughly corresponds to the 95\% Encircled Energy Fraction \citep[EEF][]{Giacconi2002, Paolillo2017}. In the analysis that follows the 0.5-7\,keV threshold of 350 net counts is adopted as the sample selection function. This limit cuts through the black-hole mass and Eddington ratio parameter space and therefore affects the expected excess variance of the detected sources as well as that of the ensemble. It is therefore necessary to apply the count limit above to the mock AGN sample to mimic the CDFS 7\,Ms observational selection effects. The adopted method for achieving this is discussed below. 

\subsection{Modeling the CDFS selection function}

The conversion of the 0.5-7\,keV flux of mock AGN to the observed photon counts on the Chandra ACIS-I detector assumes a power-law spectral model with index $\Gamma=1.4$ that is absorbed by the Galactic hydrogen column density in the direction of the CDFS, $N_H=\rm 8.8 \times 10^{19}\, cm^{-2}$ \citep{Luo2017}. The choice of $\Gamma=1.4$ is because \cite{Luo2017} adopt this value to construct the exposure maps of the CDFS, which are used in this calculation. The net counts of a source with a given flux depend on its position within the CDFS field of view. The maximum exposure is achieved close to the centre of the field and then drops smoothly toward the  edges as a result of vignetting. Therefore, at fixed flux more counts are expected close to the CDFS centre compared to the field edges. It is possible to estimate the CDFS area over which a source with a given flux has more than 350 net counts within an aperture that includes 95\% of the source photons. The fraction of this area relative to the total of the CDFS field provides a measure of the probability that sources with the flux in question are included in the  \cite{Paolillo2017} variability sample, i.e. the observational selection function. Using the CDFS 0.5-7\,keV exposure map\footnote{https://personal.psu.edu/wnb3/cdfs/cdfs-chandra.html} and the spectral model above the expected net counts within the 95\% EEF aperture is

\begin{equation}
C = f_X({\rm 0.5-7\,keV}) \cdot t \cdot ECF \cdot 0.95 ,
\end{equation}

\noindent where $t$ represents the distribution of the exposure-map pixel values, $f_X(\rm 0.5-7\,keV)$ is the energy flux in the 0.5-7\,keV band and the $ECF$ is the energy to photon-flux conversion factor. For the adopted spectral model $ECF=3.23\times10^{-9}$. The fraction of the exposure-map pixels that yield $C>350$ measure the CDFS fractional area within which a source with $f_X(\rm 0.5-7\,keV)$ has sufficient counts to be included in the variability sample of \cite{Paolillo2017}. This fraction is plotted as a function of $f_X(\rm 0.5-7\,keV)$ in Figure \ref{fig:selfun}. This curve is used to assign weights to each source in the mock catalogue and generate samples that match the \cite{Paolillo2017} selection. 


Next we assess the ability of the selection function of Figure \ref{fig:selfun} to reproduce the basic observational properties of the variability sample of CDFS AGN used in our analysis. Figure \ref{fig:lgxz2D} plots the distribution on the $L_X-z$ plane of mock AGN in the redshift interval $z=0.4-4.0$ after filtering with the selection function curve of Figure \ref{fig:selfun}. The 7\,Ms CDFS AGN \citep{Luo2017} in the same redshift range and with full-band net counts $>350$ are also plotted in Figure \ref{fig:lgxz2D} for comparison. Overall there is fair overlap in the distribution of mock and real AGN on the  $L_X-z$ parameter space. This suggests that the selection function curve of  Figure \ref{fig:selfun} provides a reasonable representation of the observational selection effects of the 7\,Ms CDFS field. This is further explored in Figures \ref{fig:histz} and \ref{fig:histlglx} that compare the redshift and luminosity distributions of mock AGN with the CDFS observations. The observed redshift peaks in Figure \ref{fig:histz} trace the substruture of the cosmic web along the CDFS line of sight, which is absent from the model. The observations also find a lower fraction of AGN in the interval $z=0.5-1.5$ compared to the model prediction. Poisson uncertainties and cosmic variance are likely responsible for this difference. Nevertheless, the model tracks reasonably well the high-redshift tail of the observations. In Figure \ref{fig:histlglx}  there is evidence for an excess of luminous AGN in the observations compared to the model predictions. This is largely because of the differences in the redshift distribution of the  model and observations in Figure \ref{fig:histz}. 
The evidence above shows that the selection function curve of  Figure \ref{fig:selfun} reproduces at least to the first approximation the observational biases of the 7\,Ms CDFS sample used by \cite{Paolillo2017}.

\subsection{Constructing the CDFS variability model}\label{sec:model_summary}

The galaxy stellar-mass function of Section \ref{sec:SMF} is used to generate a sample of galaxies in the redshift interval $z=0.4-4$ that corresponds to the  \cite{Paolillo2017}  CDFS variability sample. These are assigned specific accretion rates, X-ray luminosities and hydrogen column densities as explained in Section \ref{sec:SARD}. X-ray fluxes in the 0.5-7\,keV band are also estimated at this stage. The mock AGN are seeded with black holes using the scaling relations of Section \ref{sec:MSBH}. Eddington ratios are also estimated for individual systems. For each mock AGN the four PSD models of Section \ref{sec:psd} are integrated between the lowest and highest rest-frame frequencies sampled by the 7\,Ms CDFS light-curves as defined by Equations \ref{eq:numin} and \ref{eq:numax} ($\Delta t^{tobs}_{max}=6205$\,days and $\Delta t^{tobs}_{max}=0.25$\,days). The PSD integral yields $\sigma^2_{mod}$ for each mock AGN, which is then converted to $\sigma^2_{obs}$ via Equation \ref{eq:sampling}. In the latter calculation we adopt $C=1.3$ and $\beta=1.1$, which are appropriate for the sampling pattern of the CDFS 7Ms light curves. The selection function curve of Figure \ref{fig:selfun} is used to assign weights to each mock AGN. The ensemble excess variance of the model AGN population within X-ray luminosity bins is the weighted average of the individual $\sigma^2_{obs}$.

\begin{figure}
\begin{center}
\includegraphics[width=0.9\columnwidth]{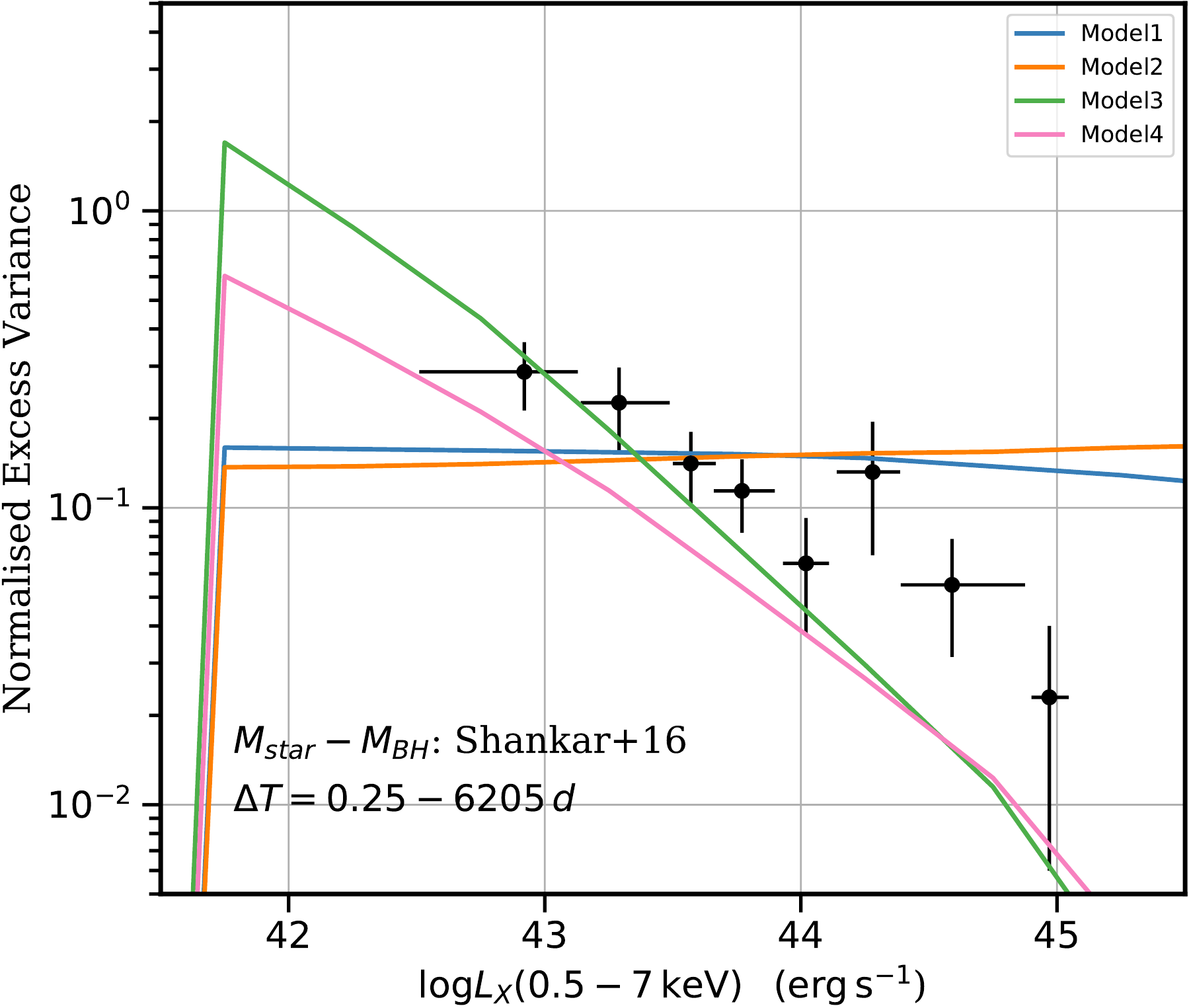}
\end{center}
\caption{Normalised excess variance of AGN as a function of X-ray luminosity in the rest-frame 0.5-7\,keV energy band. The data-points  are the measurements of the ensemble variance of the CDFS AGN in the redshift range $z=0.4-4$ presented by \protect\cite{Paolillo2017} for the longest timescale probed by these observations, 17\,yr. The curves correspond to the empirical model presented in this paper for the four different parametrisations of the adopted PSD as indicated in the legend.  These curves are based on the \protect\cite{Shankar2016} "unbiased" or "intrinsic" stellar-mass vs black-hole mass scaling relation given by Equations \ref{eq:MstarMBH_int} and \ref{eq:MstarMBH_sc}. 
}\label{fig:res_sha_stack}
\end{figure}

\begin{figure}
\begin{center}
\includegraphics[width=0.9\columnwidth]{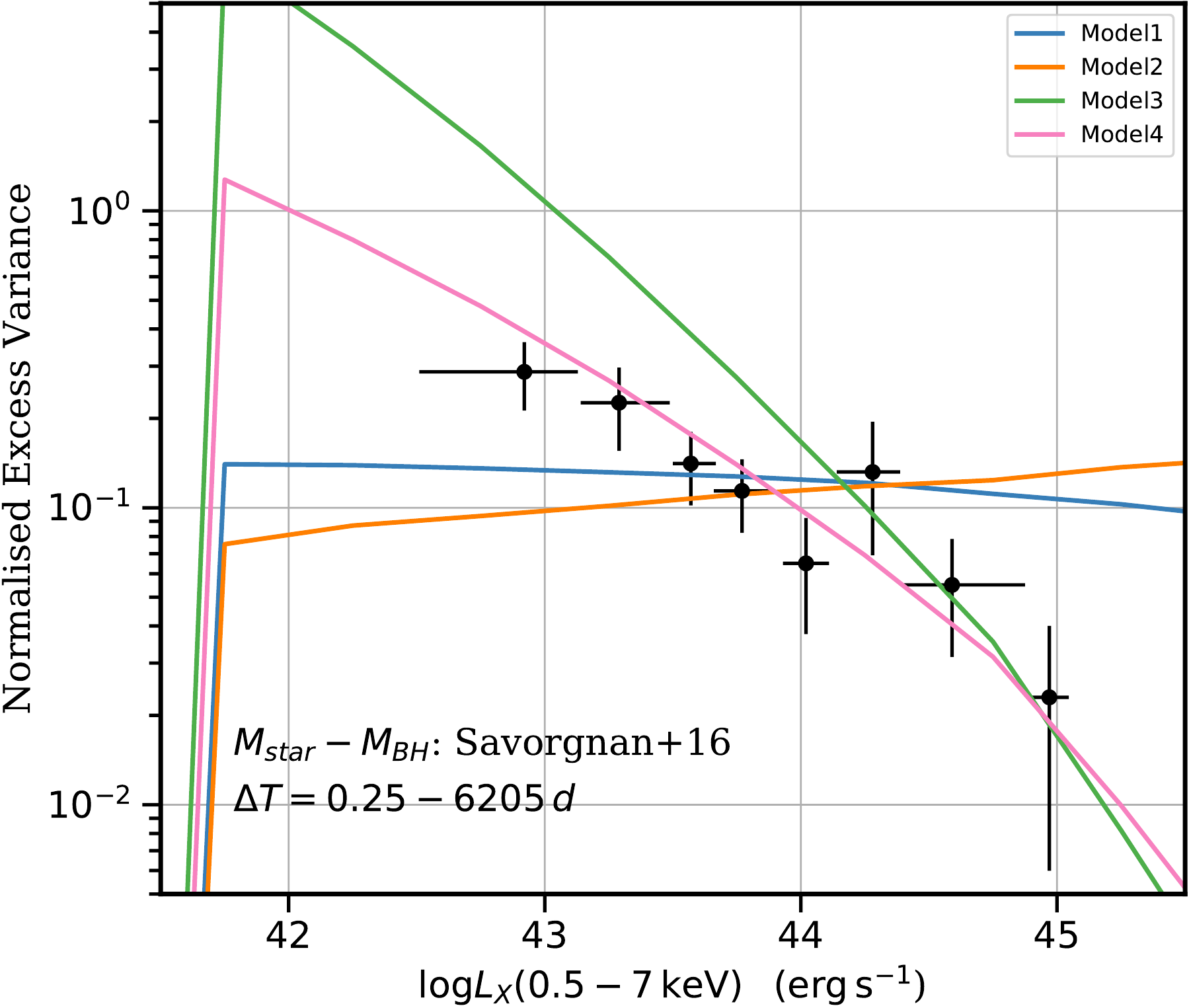}
\end{center}
\caption{Same as Figure \ref{fig:res_sha_stack} with the model curves constructed assuming the  \protect\cite{Savorgnan2016} dynamical stellar-mass vs black-hole mass scaling relation given by Equation \ref{eq:MstarMBH_dyn}.}\label{fig:res_sav_stack}
\end{figure}

\begin{figure}
\begin{center}
\includegraphics[width=0.9\columnwidth]{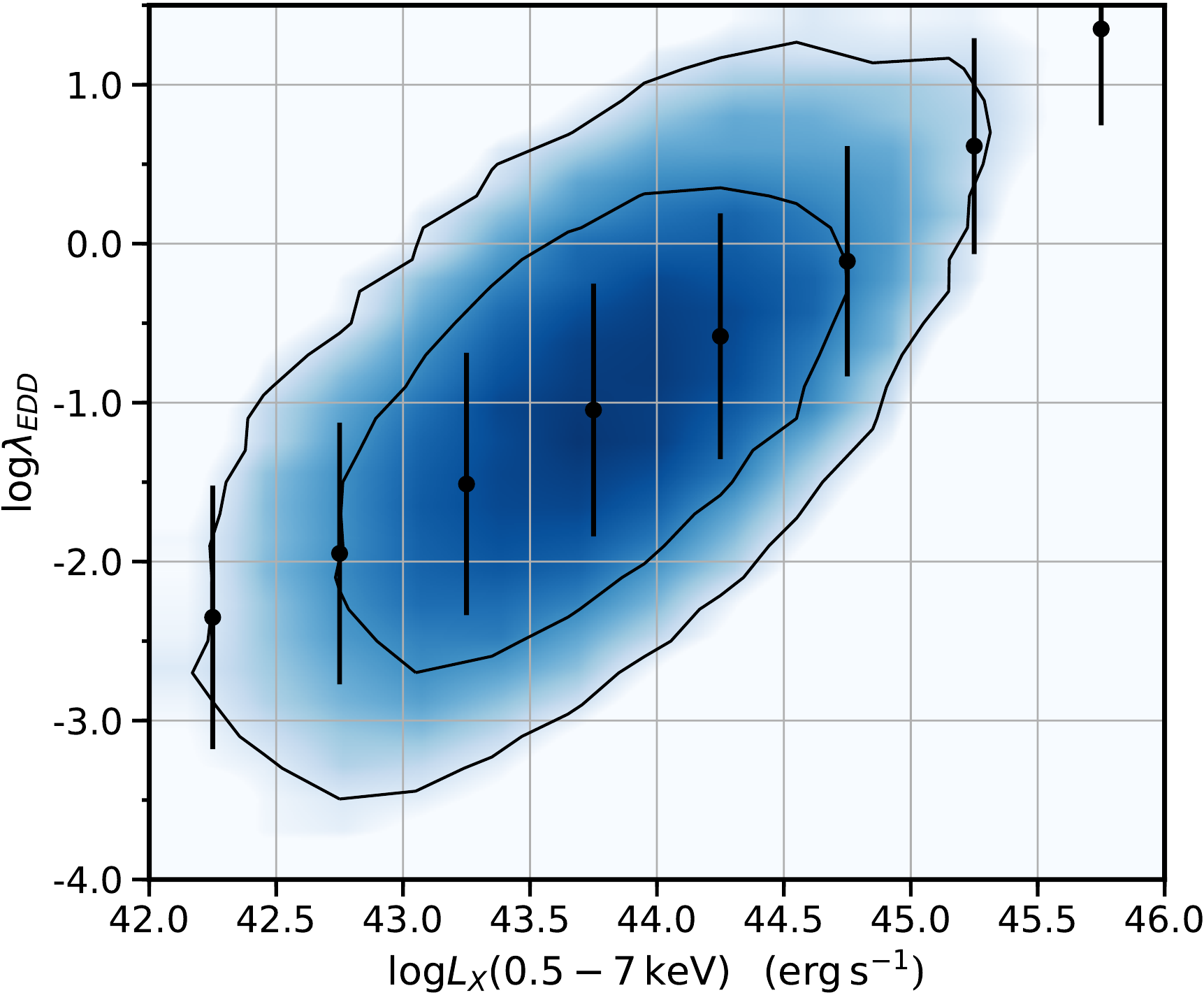}
\end{center}
\caption{Eddington ratio as a function of the X-ray luminosity in the 0.5-7\,keV band for the mock AGN sample that has been filtered through the 7\,Ms-CDFS selection function of Fig. \ref{fig:selfun}.  The contours and blue-shaded regions show the mock AGN distribution.  Darker  colours  correspond  to  a  higher density of sources. The contours enclose 68 and 95 per cent of  the  population. The datapoints and errorbars show the mean and standard deviation of the Eddington-ratio distribution in different luminosity bins. The Eddington ratio of mock AGN is estimated using the \protect\cite{Savorgnan2016} black-hole mass vs stellar mass scaling relation.}\label{fig:lxledd}
\end{figure}

\begin{figure}
\begin{center}
\includegraphics[width=0.9\columnwidth]{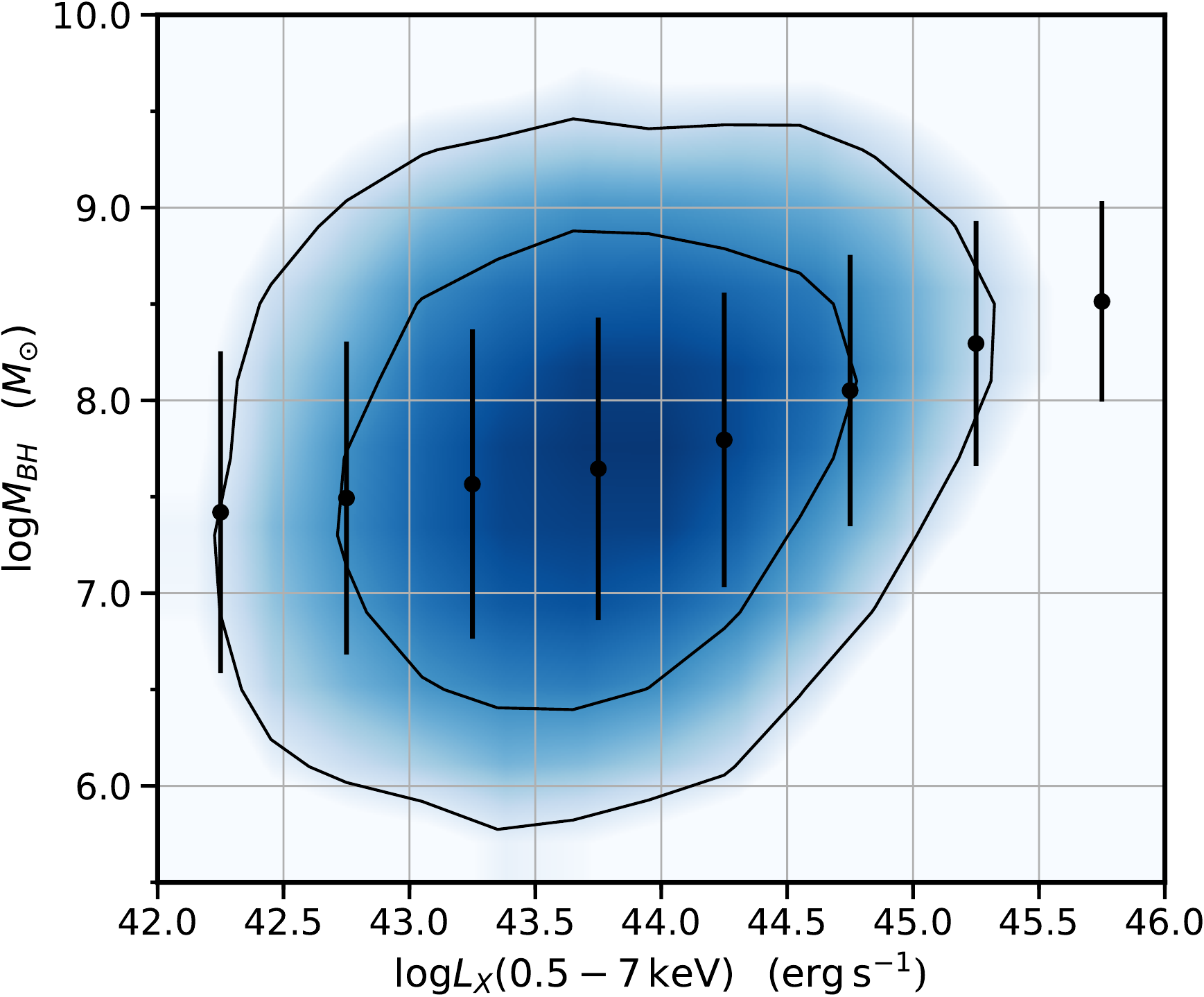}
\end{center}
\caption{Black-hole mass as a function of the X-ray luminosity in the 0.5-7\,keV band for the mock AGN sample that has been filtered through the 7\,Ms-CDFS selection function of Fig. \ref{fig:selfun}.  The contours and blue-shaded regions show the mock AGN distribution.  Darker  colours  correspond  to  a  higher density of sources. The contours enclose 68 and 95 per cent of  the  population. The datapoints and errorbars show the mean and standard deviation of the black-hole mass distribution in different luminosity bins. The black-hole mass of mock AGN is estimated using the \protect\cite{Savorgnan2016} black-hole mass vs stellar mass scaling relation.}\label{fig:lxbh}
\end{figure}

\section{Results}

\subsection{Model vs Observations}

The comparison of the CDFS ensemble variance observations with the model predictions is shown in Figures \ref{fig:res_sha_stack}  and \ref{fig:res_sav_stack} for the \cite{Shankar2017} and \cite{Savorgnan2016} black-hole mass vs stellar mass scaling relations respectively.  The curves shown in these figures correspond to the four PSD models of Section \ref{sec:psd}. They predict very different relations between ensemble excess variance and luminosity. The PSD models 1 and 2 predict flat relations, whereas in models 3 and 4 the excess variance decreases with increasing luminosity. This is a result of the dependence of the PSD amplitude on Eddington ratio in the latter group of models. Figure \ref{fig:lxledd} shows that the mean Eddington ratio of mock AGN increases with increasing X-ray luminosity. This translates to a lower normalisation of the corresponding PSDs (see Equation \ref{eq:model3-amp}) and hence, a lower ensemble excess variance with increasing luminosity. In contrast, the models 1, 2 are more rigid and the only variation in the excess variance of AGN is via the black-hole mass dependence of the PSD break frequency. Figure \ref{fig:lxbh} shows the distribution of the black-hole masses of mock AGN as a function of their X-ray luminosities. These two quantities are only weakly correlated in the sense that more luminous AGN are found in increasingly more massive black holes. This translates to a lower PSD break frequency with increasing luminosity. However, at fixed X-ray luminosity the median black hole mass only mildly increases with $L_X$. As a result for the variability timescales probed by the 7\,Ms CDFS observations the black-hole mass dependence of the PSD break-frequency is insufficient to produce a strong drop in the ensemble excess variance of AGN toward bright luminosities. 

The datapoints in Figures \ref{fig:res_sha_stack} and \ref{fig:res_sav_stack} show a decreasing trend with increasing luminosity and strongly favour the predictions of the PSD models 3 and 4. The flat excess variance curves produced by the models 1, 2 are inconsistent with these observations, irrespective of the black-hole mass vs stellar mass relation. This result shows that the ensemble variance of AGN populations can constrain PSD models and points to a variable PSD amplitude that is a function of the physical parameters of the accretion process. This is in agreement with the results of \cite{Paolillo2017}, who also favoured PSD models with amplitudes that depend on the accretion rate. However the modeling presented by \cite{Paolillo2017} did not include any apriori constraints on the black-hole masses of AGN, and therefore could not reject constant-amplitude PSD models at a high confidence level. Our modeling approach links AGN demographics with variability models and therefore contains necessary additional information on the black-hole mass distribution of AGN. The correlation between X-ray luminosity and black-hole mass in our model (Fig. \ref{fig:lxbh}) is flatter than that assumed by \cite{Paolillo2017} and hence, provides stronger constraints on the PSD parametrisation.   

Figures \ref{fig:res_sha_stack}  and \ref{fig:res_sav_stack} further show that the plotted model curves depend on the adopted stellar-mass vs black-hole mass scaling relation. At fixed redshift, X-ray luminosity and PSD parametrisation there are differences between the models that use the \cite{Shankar2017} and the \cite{Savorgnan2016} $M_\star-M_{\rm BH}$ correlations.  This indicates that measurements of the ensemble variability of AGN can constrain the $M_\star-M_{\rm BH}$ relation of the population. The comparison of the CDFS ensemble variability measurements with the PSD model 3, 4 predictions broadly favour a high normalisation for the stellar-mass vs black-hole mass scaling relation, similar that proposed by \cite{Savorgnan2016} based on dynamical black-hole mass estimates. 

The evidence above suggests that observations of the ensemble variance of AGN can jointly constrain PSD models and the relation between stellar and black-hole mass of the population.  Figures \ref{fig:res_sha_stack}, \ref{fig:res_sav_stack} provide important clues in this direction but do not explore the full range of model parameters that are consistent with the observations. Model inference is needed to sample the parameter space in a statistically robust way and provide confidence intervals to model parameters. This is also necessary to assign statistical significance to the results and allow the quantitative interpretation of the model parameters.

\subsection{Model-parameter inference}\label{sec:inference}

We start by adopting simple parametric models for the $M_{\star}-M_{\rm BH}$ relation and the AGN PSD. Bayesian inference is then used to fit the observations of Figure \ref{fig:res_sha_stack} and constrain the model parameters. The relation between stellar and black hole mass is parametrised as

\begin{equation}\label{eq:inference-bh}
    \log\,M_{\rm BH} = \alpha + \beta \cdot (\log\,M_{\star} - 10).
\end{equation}

\noindent The intrinsic scatter of this relation is fixed to 0.5\,dex, i.e. comparable to that inferred by \cite{Shankar2017} and \cite{Savorgnan2016} for their scaling relations. 

The adopted PSD parametrisation is based  on the Model 3 of Section \ref{sec:psd}. This is the simplest model that reproduces the observed luminosity dependence of the ensemble excess variance (see Figures  \ref{fig:res_sha_stack}  and \ref{fig:res_sav_stack}). For the current inference application both the exponent and normalisation are free parameters 

\begin{equation}\label{eq:inference-psd}
A = 2\cdot\nu_b \cdot PSD(\nu_b) = \delta \cdot\lambda_{Edd}^{\gamma}.
\end{equation}

\noindent The AGN ensemble variability model therefore has a total of 4 free parameters, two related to the  $M_{\star}-M_{\rm BH}$ relation ($\alpha$, $\beta$) and the remaining ($\gamma$, $\delta$) to the PSD model. We choose not to expand further the parameter space, e.g. by adding non-linear terms to Equation \ref{eq:inference-bh} or allowing the scatter of this relation to be a free parameter. This is because the current observational constraints, although state-of-the-art, still have relatively large uncertainties, which ultimately relate to small number statistics. 


The free parameters $\alpha$, $\beta$, $\gamma$ and $\delta$ are determined by sampling the likelihood 

\begin{equation}
    \mathcal{L} = -\frac{1}{2}\,\sum_{i} \frac{( \sigma^2_{NXV, i} - \sigma^2_{obs,i})^2}{\delta^{2}_{NXV, i}},
\end{equation}

\noindent where $\sigma^2_{NXV, i}$ is the measured ensemble normalised excess variance for the X-ray luminosity bin $i$ and $\delta^2_{NXV, i}$ is the corresponding uncertainty. The symbol $\sigma^2_{obs,i}$ is the ensemble excess variance predicted by the model for the luminosity bin $i$ (Equations \ref{eq:s2mod} and \ref{eq:sampling}). The likelihood above assumes that the excess variance measurements are normally distributed with a scatter that is represented by the corresponding uncertainty shown in Figures   \ref{fig:res_sha_stack} and \ref{fig:res_sav_stack} \citep[see][]{Allevato2013_var}. The {\sc MultiNest} multimodal nested sampling algorithm \citep{Feroz2008, Feroz2009} is used for parameter estimation. Flat priors are adopted for the model parameters $\alpha$, $\beta$ within the intervals 0--9 and 0--3, respectively. The choice of the priors for the parameters $\gamma$, $\delta$ is informed by results on the variability of local AGN presented by \cite{Ponti2012}. They studied the excess variance of their sample as function of black-hole mass and Eddington ratio and measured $\gamma=-0.8\pm0.15$ and $\delta=0.003^{+0.002}_{-0.001}$. Based on these independent observational result we choose a Gaussian prior for $\gamma$ with mean  $-0.8$ and scatter $\sigma=0.15$. The amplitude $\delta$ of the PSD can only take positive values. We therefore set a Gaussian prior for the parameter $\log\delta$ with a mean of $\log(0.003)=-2.522$ and scatter 0.20, which corresponds to the logarithmic uncertainty of the mean $1\sigma$ rms errors estimated by \cite{Ponti2012}.

\begin{figure}
\begin{center}
\includegraphics[width=0.99\columnwidth]{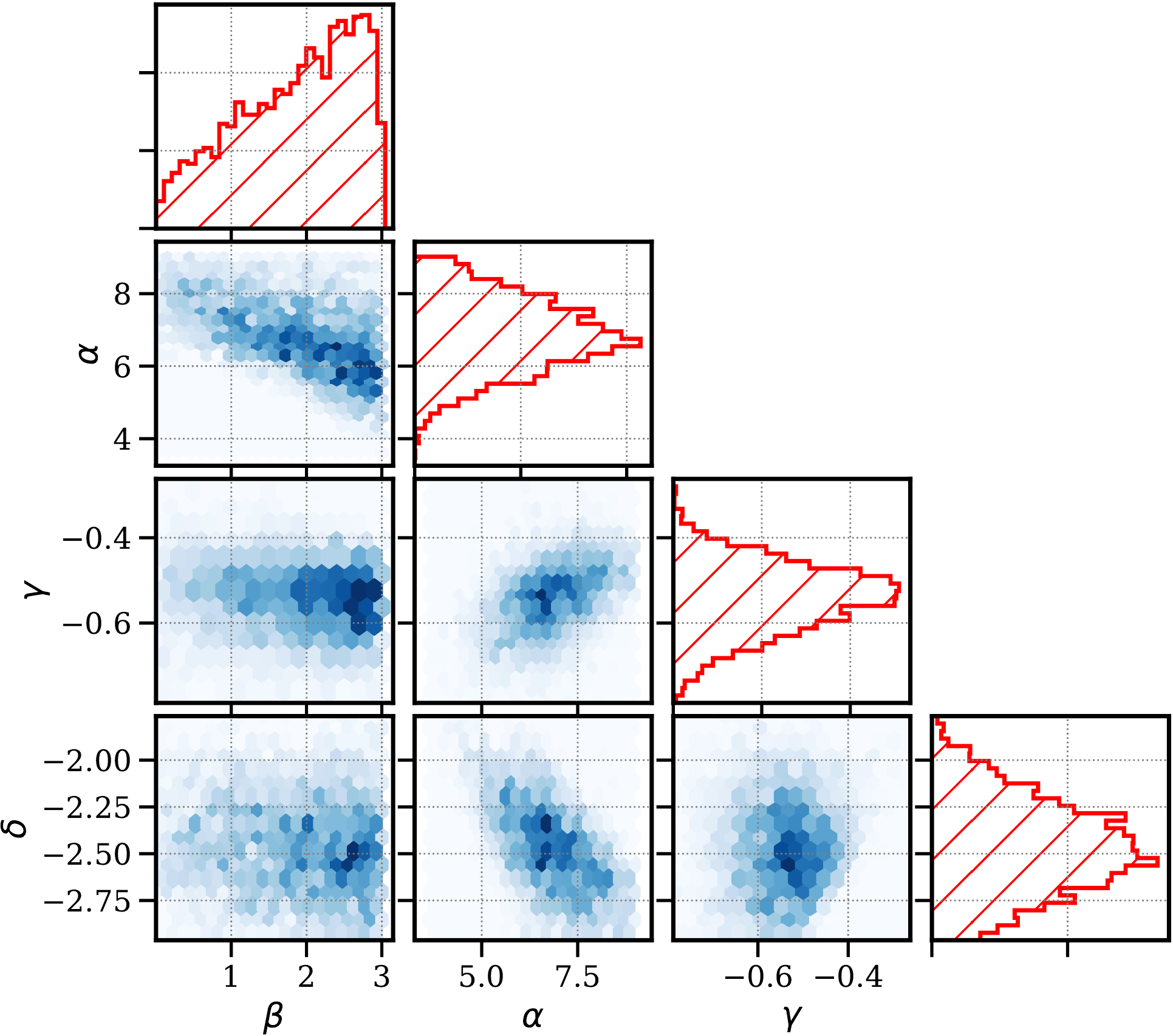}
\end{center}
\caption{Corner plots of the posterior distributions of the parameters $\alpha$, $\beta$, $\gamma$, $\delta$ of the model presented in Section \ref{sec:inference}. These plots include both 2-dimensional and 1-dimensional projections of the posterior distribution.}\label{fig:triangle}
\end{figure}

Table \ref{tab:inference} lists the inference results for the four model parameters. Figure \ref{fig:triangle} shows the corner plot of the parameter posterior distributions. There is a strong covariance between the slope, $\beta$ and normalisation, $\alpha$, of the scaling relation between black-hole and stellar mass. Also, the slope $\beta$ is largely unconstrained by the current ensemble variability observations. This is manifested by the broadness of the $\beta$ posterior distribution, which is comparable to the adopted prior for this parameter (flat between 0 and 3). Aliases also exist between the parameters $\alpha$ and $\gamma$, in the sense that lower $\gamma$ values broadly correspond to  lower normalisations of the $M_{\star}-M_{\rm BH}$ relation. Based on the posterior distributions of Figure \ref{fig:triangle} we find $\gamma=-0.54^{+0.07}_{-0.09}$ (Table \ref{tab:inference}), i.e. shallower than the prior $\gamma=-0.8\pm0.15$, which is estimated by \cite{Ponti2012}  for local Seyferts. This shows that observations of the ensemble variance of AGN in deep survey fields provide additional information on the PSD of the population. 

Also shown in Figure \ref{fig:inference} are the constraints on the $M_{\star}-M_{\rm BH}$ relation using the posterior distributions of the model parameters $\alpha$, $\beta$ of Equation \ref{eq:inference-bh}. The joint fit to the PSD and the $M_{\star}-M_{\rm BH}$ models produces results that are consistent with the scaling relation of \cite{Savorgnan2016}. Nevertheless the inferred 68\% confidence region around the median is broad and therefore the \cite{Shankar2017} relation cannot be excluded at a high confidence level. Improved ensemble variability measurements have the potential to provide better constraints on the $M_{\star}-M_{\rm BH}$ scaling relation of the AGN population and test different parametrisations proposed in the literature.  For completeness Figure \ref{fig:inference} also plots the projection of the model onto the observed space of the ensemble excess variance and X-ray luminosity. 

The inferred parameters on the $M_{\star}-M_{\rm BH}$ relation in Table \ref{tab:inference} are sensitive to the assumption that the stellar mass is a proxy to the bulge mass of galaxies. We explore the impact of this effect on the results by setting $M_{bulge}=0.5 \cdot M_\star$ for all mock AGN, i.e. similar to the bulge--to--total-mass ratio of Sb/Sbc-type galaxies \citep[e.g.][]{Fukugita1998,  Oohama2009}. The stellar mass, $M_\star$, in Equation \ref{eq:inference-bh} is then substituted by the $M_{bulge}$. This results to lower black-hole masses and an overall lower $\sigma_{obs}$ for the model AGN. Fitting the observations of Figure \ref{fig:inference} therefore requires a higher normalisation of the $M_{\star}-M_{\rm BH}$  relation by about 0.5\,dex. Observational constraints on the average $M_{bulge}/M_\star$ ratio as a function of  stellar-mass and redshift could help mitigate this systematic by including statistical corrections into the modeling.

\begin{table}
\begin{tabular}{cc cc}
\hline
parameter &  median value & prior & prior \\
          & and error     &  type & parameters\\
(1)     & (2)             & (3)   & (4) \\
\hline
$\log \alpha$ & $6.8^{+1.0}_{-1.1}$  & flat & [0--9] \\
$\beta$  & $2.0^{+0.7}_{-1.0}$  & flat & [0--3] \\
$\gamma$ & $-0.54^{+0.07}_{-0.09}$ & normal & $\mu = -0.8$, $\sigma = 0.15$ \\
$\log \delta$ & $-2.5^{+0.3}_{-0.3}$ & normal & $\mu = -2.522$, $\sigma = 0.2$ \\
\hline
     \end{tabular}
\caption{Bayesian inference results for the parameters $\alpha$\, $\beta$, $\gamma$ and $\delta$ of Equations \ref{eq:inference-bh}, \ref{eq:inference-psd}. The columns are (i) parameter name, (ii) the median value of the parameter and the corresponding uncertainty measured as the interval around the median that includes 68\% of the probability density, (iii) the type of prior, flat of Gaussian, used in the inference and (iv) the prior range defined as the interval of the flat prior and the mean ($\mu$), scatter ($\sigma$) of the Gaussian prior.}\label{tab:inference}
\end{table}

\subsection{Future prospects}

Current measurements of the ensemble variance of AGN are limited by the size of the available samples. The number of extragalactic X-ray fields with sufficient number of multi-epoch observations is small. The eROSITA 4-year All Sky Survey  \citep{Predehl2020} will change this by providing 4-year light curves with a 6-month cadence over a $4\pi$ solid angle. The large number of AGN in this survey combined  with the modeling methodology described in this paper has the potential to provide unique joint constraints on the PSD and  $M_{\star}-M_{\rm BH}$ relation as a function of redshift. The ensemble excess variance predicted by the model can generically be expressed as

\begin{equation}
\sigma_{obs}^2 =f(L_X, M_{\star}),
\end{equation}

\noindent where we have assumed that the $L_X$ is a proxy of the bolometric luminosity and the $M_\star$ is a measure of the black-hole mass. The function $f(L_X, M_{\star})$ encapsulates the dependence of the ensemble excess variance on the PSD model, the form of the $M_\star - M_{BH}$ relation and the parametrisation of the observational selection effects. The equation above shows that the ($L_X$, $M_\star$) plane is the natural choice of parameter space for future ensemble variance measurements that are not limited by small number statitistics. Providing measurements of the  $\sigma_{NXV}^2$ in bins of stellar mass and accretion luminosity has the potential to minimise aliases between parameters of interest and improve the robustness of the results. Additionally, splitting samples into distinct redshift intervals can provide a handle on the cosmic evolution of e.g. the $M_{\star}-M_{\rm BH}$ relation.

\section{Discussion and Conclusions}

A new model for the ensemble X-ray variability of AGN selected in extragalactic surveys is presented. It is developed upon empirical relations and is designed to account for observational selection effects, such as flux limits. The starting point of the model are observational measurements of the incidence of X-ray AGN among galaxies. These are combined with analytic expressions for the variability PSD of AGN and the $M_{\star}-M_{\rm BH}$ scaling relation to make predictions on the ensemble excess variance as a function of observables, such as accretion luminosity and redshift.

Comparison with observational measurements of the AGN ensemble excess variance \citep{Paolillo2017} shows that the empirical model has predictive power and can constrain parameters related to e.g. the AGN PSD and the $M_{\star}-M_{\rm BH}$ relation. Our modeling favours PSD models approximated by a double power-law with amplitude that depends on the Eddington ratio and/or the black-hole mass of the accreting system. Parametrisations of the PSD in which only the break frequency of the double power-law depends on the physical properties of the AGN are unable to reproduce the observed decreasing trend of the ensemble excess variance with increasing X-ray luminosity. These constraints can feedback to studies of the light-curves of individual local AGN \citep[e.g.][]{Ponti2012} to provide independent information on the PSD parameterisation.  Similar conclusions, but at lower significance, are presented by \cite{Paolillo2017}.

\begin{figure}
\begin{center}
\includegraphics[width=0.99\columnwidth]{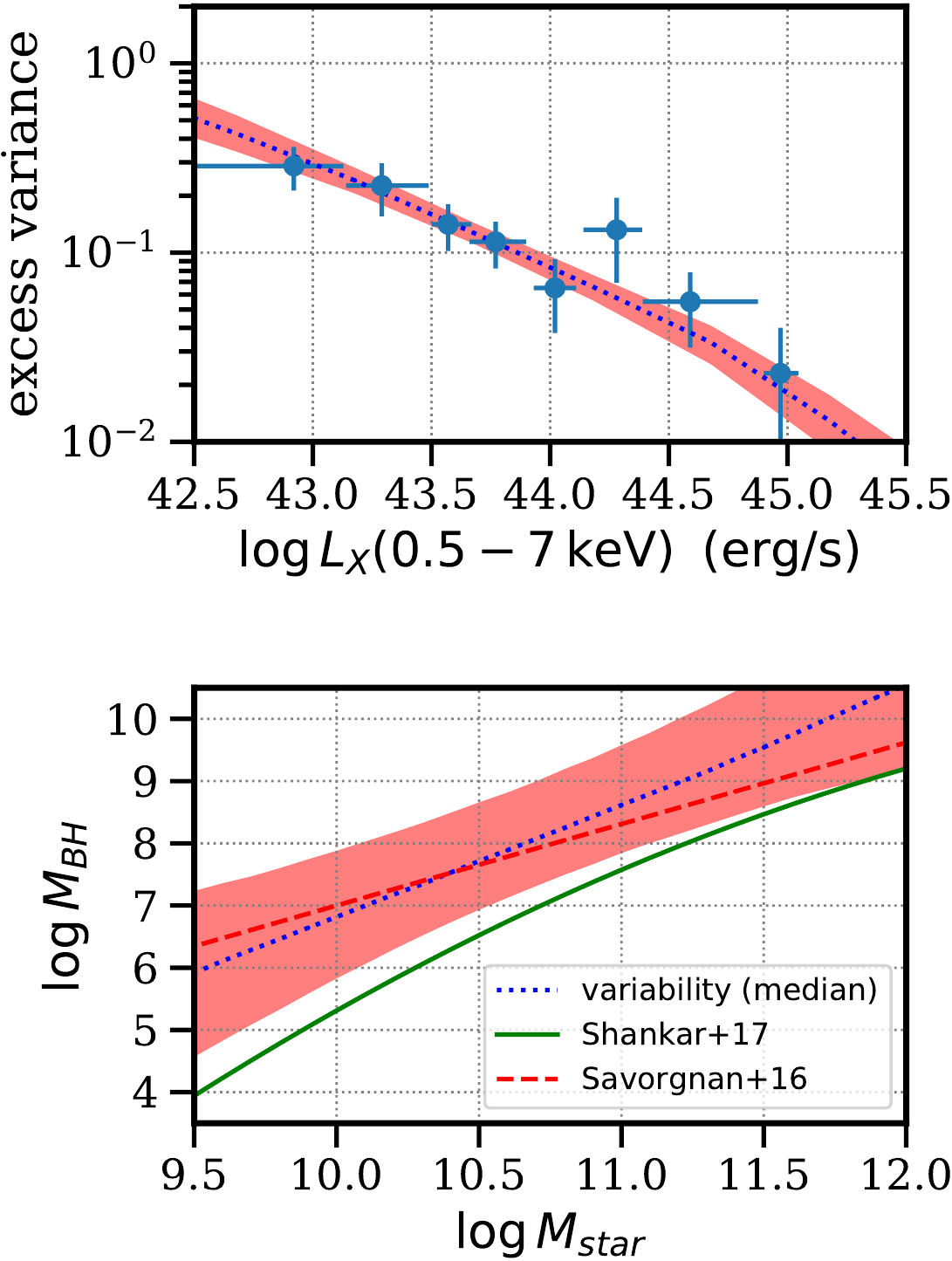}
\end{center}
\caption{The top panel shows the projection of the model of Section \ref{sec:inference} onto the $\sigma^2_{NXV}$ vs $L_X$ space. The data-points  are the measurements of the ensemble variance of the CDFS AGN in the redshift range $z=0.4-4$ presented by \protect\cite{Paolillo2017} for the longest timescale probed by these observations, 17\,yr. The pink-shaded region is the model prediction using the posterior distributions shown in Fig. \ref{fig:triangle}. The extent of the pink region corresponds to the 68\% confidence interval around the median. The bottom panel shows the  $M_{\star}-M_{\rm BH}$ relation based on the posterior distribution of the parameters $\alpha$, $\beta$ (see Equation \ref{eq:inference-bh}). The blue dashed line shows the median at each $M_{\star}$ bin. The pink-shaded region corresponds to the 68\% confidence interval around the median. Also shown for comparison are the \protect\cite{Savorgnan2016} dynamical stellar-mass vs black-hole mass relation (red dashed line) and the \protect\cite{Shankar2016} "unbiased" or "intrinsic" scaling relation (green solid curve).}\label{fig:inference}
\end{figure}

An interesting feature of the empirical model developed in this paper is that it has the potential to constrain the black-hole mass vs stellar mass relation of AGN samples based on measurements of their mean variability properties.  This opens the possibility to constrain the redshift evolution of this relation and to complement studies that use spectral methods to directly measure black-hole masses of QSO samples and relate them to the properties of their hosts \citep[e.g.][]{Treu2004, Jahnke2009, Shen2015, Sexton2019, Li2021}. We find that the ensemble variability observations of  \cite{Paolillo2017} favour a normalisation of the $M_{\star}-M_{\rm BH}$ relation similar to that measured in the local Universe for dormant black holes in quiescent spheroids \citep[e.g.][]{Kormendy_Ho2013}. Nevertheless the variability constraints shown in Figure \ref{fig:inference} cannot reject at a high confidence level a lower normalisations of the $M_{\star}-M_{\rm BH}$ relation proposed by \cite{Shankar2017}. A substantial increase in the size of X-ray AGN samples with repeat observations is needed to improve current constraints and explore the redshift evolution of black-hole/host-galaxy scaling relations.  The eROSITA All Sky Survey has the potential to deliver such a dataset. 

\section{Data and Code Availability}

The code and data used in this paper are available at
\url{https://github.com/ageorgakakis/EnsembleVariability} and
\url{https://zenodo.org/record/4725121}.

\bibliographystyle{mnras}
\bibliography{mybib} 

\appendix 

\bsp	
\label{lastpage}
\end{document}